\documentclass[11pt]{article}
\usepackage[totalwidth=460truept,totalheight=600truept]{geometry}
\usepackage{epsfig,latexsym}

\global\arraycolsep=1truept
\input epsf
\input{tcilatex}
\begin{document}

\null

\font\cmss=cmss10 \font\cmsss=cmss10 at 7pt \hfill \hfill IFUP-TH/04-07


\vspace{10pt}\vskip 1.5truecm

\begin{center}
{\Large \textbf{A\ NOTE\ ON\ THE\ DIMENSIONAL REGULARIZATION \\[0pt]
OF THE STANDARD MODEL COUPLED WITH QUANTUM\ GRAVITY}}

\bigskip \vskip 1truecm

\textsl{Damiano Anselmi}

\textit{Dipartimento di Fisica ``E. Fermi'', Universit\`{a} di Pisa, and INFN%
}
\end{center}

\vskip 2truecm

\begin{center}
\textbf{Abstract}
\end{center}

\bigskip

{\small In flat space, }$\gamma _{5}${\small \ and the epsilon tensor break
the dimensionally continued Lorentz symmetry, but 
propagators have fully Lorentz invariant denominators. When the Standard
Model is coupled with quantum gravity }$\gamma _{5}${\small \ breaks the
continued local Lorentz symmetry. I show how to deform the Einstein
lagrangian and gauge-fix the residual local Lorentz symmetry so that the
propagators of the graviton, the ghosts and the BRST auxiliary fields have
fully Lorentz invariant denominators. This makes the calculation of Feynman
diagrams more efficient.}

\vskip 1truecm

\vfill\eject

The dimensional-regularization technique \cite{bollini,thooft} is the most
efficient technique for the calculation of Feynman diagrams in quantum field
theory. Its main virtue is that it is manifestly gauge invariant, when gauge
bosons couple to fermions in a chiral invariant way. When gauge bosons
couple to chiral currents, gauge anomalies can be generated. If the gauge
anomalies vanish at one-loop, as in the Standard Model, then, by the
Adler-Bardeen theorem \cite{adler}, there exists a subtraction scheme where
they vanish at each order of perturbation theory. This ensures internal
consistency.

The definition of $\gamma _{5}$ in dimensional regularization \cite
{thooft,maison,collins} breaks the Lorentz symmetry in the dimensionally
continued spacetime. The calculation of Feynman diagrams in parity violating
theories is still efficient, because the continued Lorentz symmetry is not
broken in the denominators of propagators, but only in vertices and
numerators of propagators. Using appropriate projectors, a Feynman integral
can be decomposed in a basis of scalar and fully Lorentz invariant
integrals. The complication introduced by $\gamma _{5}$ is only algebraic
and a computer can easily deal with it. Calculations have the same
conceptual difficulty than in the parity invariant theories.

When the Standard Model is coupled with quantum gravity, $\gamma _{5}$
breaks the dimensionally continued \textit{local} Lorentz symmetry. It is
less obvious how to break the continued local Lorentz symmetry and maintain
efficiency in the calculation of Feynman diagrams. In this paper I show how
this can be done.

In the vielbein formalism the Einstein lagrangian 
\begin{equation}
\mathcal{L}=\frac{1}{2\kappa ^{2}}\sqrt{g}R  \label{LE}
\end{equation}
is more symmetric than the complete theory and must be supplemented with
appropriate evanescent terms. It is natural to look for an arrangement of
the regularization technique such that the denominators of propagators are
fully Lorentz invariant. The symmetric gauge is not allowed, but a
derivative gauge-fixing for the residual Lorentz symmetry, combined with a
certain trick for the BRST\ auxiliary fields, do the job. The prescription
of this paper works in arbitrary dimensions.

I work in the Euclidean framework. The conversion to the Minkowskian
framework is straightforward. I denote the physical spacetime dimension with 
$D$ and the continued dimension with $d=D-\varepsilon $. The Einstein action
(\ref{LE}) is $SO(d)$ invariant, while the complete theory is only assumed
to be $SO(D)\otimes SO(-\varepsilon )$ invariant. Since no confusion can
arise, the Euclidean $SO(...)$ symmetries will be called ``Lorentz''
symmetries.

Before dealing with the $SO(d)$ breaking theories it is instructive to
reformulate the regularization of pure gravity and gravity coupled with
parity invariant matter.\ Here no $SO(d)$ breaking occurs, yet it is
convenient to use a derivative gauge fixing for the Lorentz symmetry (rather
than the symmetric gauge), because it admits a straightforward
generalization to the case of gravity coupled with parity violating matter.

\bigskip

The vielbein is defined as usual in $d$ dimensions. The curved space
conventions for torsion, curvatures, covariant derivatives and connections
are 
\begin{eqnarray}
\qquad \qquad \mathcal{D}e^{a} &=&\mathrm{d}e^{a}-\omega ^{ab}e^{b}=0,\qquad
\qquad \Gamma _{\mu \nu }^{\rho }=e^{\rho a}\partial _{\mu }e_{\nu
}^{a}+\omega _{\mu }^{ab}e_{\nu }^{a}e^{\rho b},  \nonumber \\
\omega _{\mu }^{ab} &=&\frac{1}{2}\left( \partial _{\mu }e_{\nu
}^{a}-\partial _{\nu }e_{\mu }^{a}\right) e^{\nu b}-\frac{1}{2}\left(
\partial _{\mu }e_{\nu }^{b}-\partial _{\nu }e_{\mu }^{b}\right) e^{\nu a}+%
\frac{1}{2}g_{\mu \nu }\left( e_{{}}^{\rho b}\partial _{\rho }e^{\nu
a}-e^{\rho a}\partial _{\rho }e^{\nu b}\right) ,  \nonumber \\
R^{ab} &=&\frac{1}{2}R_{\mu \nu }^{ab}\mathrm{d}x^{\mu }\mathrm{d}x^{\nu }=%
\mathrm{d}\omega ^{ab}-\omega ^{ac}\omega ^{cb},\qquad {R^{\mu }}_{\nu \rho
\sigma }=\partial _{\sigma }\Gamma _{\nu \rho }^{\mu }-\partial _{\rho
}\Gamma _{\nu \sigma }^{\mu }-\Gamma _{\nu \sigma }^{\lambda }\Gamma
_{\lambda \rho }^{\mu }+\Gamma _{\nu \rho }^{\lambda }\Gamma _{\lambda
\sigma }^{\mu },  \nonumber \\
\mathcal{D}_{\mu }\psi _{i} &=&\partial _{\mu }\psi _{i}-\frac{i}{4}\omega
_{\mu }^{ab}\sigma ^{ab}\psi _{i}+iA_{\mu }\psi _{i}+A_{\mu
}^{a}T_{ij}^{a}\psi _{j}+....  \label{conve}
\end{eqnarray}
The Ricci tensor and the scalar curvature are defined as $R_{\mu \nu
}=R_{\mu \rho }^{ab}e^{\rho b}e_{\nu }^{a}$, $R=R_{\mu \nu }g^{\mu \nu },$
where of course $g_{\mu \nu }=e_{\mu }^{a}e_{\nu }^{a}$. The BRST
transformations are 
\begin{eqnarray}
se_{\mu }^{a} &=&-e_{\rho }^{a}\partial _{\mu }C^{\rho }-C^{\rho }\partial
_{\rho }e_{\mu }^{a}-C^{ab}e_{\mu }^{b},\qquad  \nonumber \\
sC^{\rho } &=&-C^{\sigma }\partial _{\sigma }C^{\rho },\qquad s\overline{C}%
_{\mu }=B_{\mu },\qquad sB_{\mu }=0,  \nonumber \\
sC^{ab} &=&-C^{ac}C^{cb}-C^{\rho }\partial _{\rho }C^{ab},  \label{brs1} \\
s\overline{C}^{ab} &=&B^{ab}-C^{ac}\overline{C}^{cb}-C^{bc}\overline{C}%
^{ac}-C^{\rho }\partial _{\rho }\overline{C}^{ab},  \nonumber \\
sB^{ab} &=&-C^{ac}B^{cb}-C^{bc}B^{ac}-C^{\rho }\partial _{\rho }B^{ab} 
\nonumber
\end{eqnarray}
Here $C^{\mu }$, $\overline{C}_{\mu }$, $B_{\mu }$ are the ghosts,
antighosts and auxiliary fields of diffeomorphisms, while $C^{ab}$, $%
\overline{C}^{ab}$, $B^{ab}$ are those of the $SO(d)$ local Lorentz symmetry.

Perturbation theory around flat space is defined as 
\begin{equation}
e_{\mu }^{a}=\delta _{\mu }^{a}+\widetilde{\phi }_{\mu }^{a}.  \label{expa}
\end{equation}
The matrix $\widetilde{\phi }$ is decomposed into its symmetric and
antisymmetric components $\phi $ and $\phi ^{\prime }$, respectively: 
\[
\widetilde{\phi }_{ab}^{{}}=\delta _{ac}\widetilde{\phi }_{\mu }^{c}\delta
_{b}^{\mu }=\phi _{ab}+\phi _{ab}^{\prime }. 
\]
Diffeomorphisms can be gauge-fixed with the common Lorentz-invariant gauge
functions 
\begin{equation}
\mathcal{G}^{\mu }\equiv \partial _{\nu }\left( \sqrt{g}g_{{}}^{\mu \nu
}\right) .  \label{gaugeD}
\end{equation}
The gauge-fixing and ghost lagrangians are the BRST\ variation of 
\[
\overline{C}_{\mu }\left( \mathcal{G}^{\mu }-\frac{\lambda }{2}B_{\mu
}\right) . 
\]
Integrating the auxiliary field $B_{\mu }$ out, we find the familiar
expressions 
\begin{equation}
\mathcal{L}_{\text{gf}}^{\text{diff}}=\frac{1}{2\lambda }(\mathcal{G}^{\mu
})^{2},\qquad \mathcal{L}_{\text{ghost}}^{\text{diff}}=\sqrt{g}\partial
_{\nu }\overline{C}_{\mu }\left( \mathcal{D}^{\mu }C^{\nu }+\mathcal{D}^{\nu
}C^{\mu }-g^{\mu \nu }\mathcal{D}_{\alpha }C^{\alpha }\right) .
\label{ghostu}
\end{equation}

\bigskip

\textbf{Lorentz gauge-fixing for }$SO(d)$ \textbf{invariant theories}. The
most popular gauge-fixing of the Lorentz symmetry is $\phi _{ab}^{\prime }=0$
(symmetric gauge). This is not very convenient for the generalization to
parity violating matter. For reasons that will become clear later, I fix the
Lorentz symmetry by means of the gauge-fixing functions 
\begin{equation}
\mathcal{G}_{L}^{ab}=\mathcal{D}^{\mu }\omega _{\mu }^{ab}=\frac{1}{\sqrt{g}}%
\partial _{\mu }\left( \sqrt{g}g^{\mu \nu }\omega _{\nu }^{ab}\right) .
\label{gaugeL}
\end{equation}
These functions are scalars under diffeomorphisms, so the ghost lagrangian
is already diagonalized. Observe that the gauge-fixing (\ref{gaugeL}) is
higher-derivative. To avoid propagators behaving like $1/p^{4}$ in the
infrared, the ``auxiliary'' fields $B^{ab}$ have to be inserted in an
unconventional derivative way. This is legitimate, because the fields $%
B^{ab} $ are anyway BRST-exact. Precisely, the gauge-fixing and ghost
lagrangian of the Lorentz symmetry are the BRST variation of 
\[
-\sqrt{g}\overline{C}^{ab}\left( \frac{\xi }{2}\mathcal{D}^{2}B^{ab}+%
\mathcal{G}_{L}^{ab}\right) , 
\]
where $\xi $ is an arbitrary gauge-fixing parameter (that can be set to zero
in the ``Landau'' gauge) and $\mathcal{D}^{2}=\mathcal{D}^{\mu }\mathcal{D}%
_{\mu }$: 
\begin{equation}
\mathcal{L}_{\text{gf}}^{L}=-\sqrt{g}\left( \frac{\xi }{2}B^{ab}\mathcal{D}%
^{2}B^{ab}+B^{ab}\mathcal{G}_{L}^{ab}\right) ,\qquad \qquad \mathcal{L}_{%
\text{ghost}}^{L}=-\sqrt{g}\overline{C}^{ab}\mathcal{D}^{\mu }\partial _{\mu
}C^{ab}.  \label{ghostv}
\end{equation}

In total, the gauge-fixed and ghost lagrangians are 
\begin{equation}
\mathcal{L}_{\text{grav}}=\mathcal{L}+\mathcal{L}_{\text{gf}}^{\text{diff}}+%
\mathcal{L}_{\text{gf}}^{L},\qquad \mathcal{L}_{\text{ghost}}=\mathcal{L}_{%
\text{ghost}}^{\text{diff}}+\mathcal{L}_{\text{ghost}}^{L}.  \label{lagrala}
\end{equation}
The ghost propagators equal the identity divided by $p^{2}$. The graviton
propagators are 
\begin{eqnarray}
\left\langle \phi _{\mu \nu }(p)~\phi _{\rho \sigma }(-p)\right\rangle _{0}
&=&\frac{\kappa ^{2}}{2p^{2}}(\delta _{\mu \rho }\delta _{\nu \sigma
}+\delta _{\mu \sigma }\delta _{\nu \rho })-\frac{\kappa ^{2}\delta _{\mu
\nu }\delta _{\rho \sigma }}{(d-2)p^{2}}+  \nonumber \\
&&+\frac{\lambda -2\kappa ^{2}}{4p^{4}}(\delta _{\mu \rho }p_{\nu }p_{\sigma
}+\delta _{\mu \sigma }p_{\nu }p_{\rho }+\delta _{\nu \rho }p_{\mu
}p_{\sigma }+\delta _{\nu \sigma }p_{\mu }p_{\rho }),  \nonumber \\
\left\langle \phi _{\mu \nu }(p)~\phi _{\rho \sigma }^{\prime
}(-p)\right\rangle _{0} &=&\frac{\lambda }{4p^{4}}(\delta _{\mu \rho }p_{\nu
}p_{\sigma }-\delta _{\mu \sigma }p_{\nu }p_{\rho }+\delta _{\nu \rho
}p_{\mu }p_{\sigma }-\delta _{\nu \sigma }p_{\mu }p_{\rho }),
\label{propagators} \\
\left\langle \phi _{\mu \nu }^{\prime }(p)~\phi _{\rho \sigma }^{\prime
}(-p)\right\rangle _{0} &=&-\frac{\xi }{2p^{2}}(\delta _{\mu \rho }\delta
_{\nu \sigma }-\delta _{\mu \sigma }\delta _{\nu \rho })+\frac{\lambda }{%
4p^{4}}(\delta _{\mu \rho }p_{\nu }p_{\sigma }-\delta _{\mu \sigma }p_{\nu
}p_{\rho }-\delta _{\nu \rho }p_{\mu }p_{\sigma }+\delta _{\nu \sigma
}p_{\mu }p_{\rho }),  \nonumber \\
\left\langle \phi _{\mu \nu }(p)~B_{\rho \sigma }(-p)\right\rangle _{0}
&=&0,\quad \left\langle \phi _{\mu \nu }^{\prime }(p)~B_{\rho \sigma
}(-p)\right\rangle _{0}=\frac{1}{2p^{2}}(\delta _{\mu \rho }\delta _{\nu
\sigma }-\delta _{\mu \sigma }\delta _{\nu \rho }),\quad \left\langle B_{\mu
\nu }(p)~B_{\rho \sigma }(-p)\right\rangle _{0}=0.  \nonumber
\end{eqnarray}

A more standard structure in (\ref{ghostv}), with non propagating auxiliary
fields $B^{ab}$, namely 
\[
\mathcal{L}_{\text{gf}}^{L}=-\sqrt{g}\left( \frac{\xi }{2}B^{ab}B^{ab}+B^{ab}%
\mathcal{G}_{L}^{ab}\right) , 
\]
can be obtained from (\ref{ghostv}) formally replacing $\xi $ with $\xi /%
\mathcal{D}^{2}$. The corresponding propagators are obtained replacing $\xi $
with $-\xi /p^{2}$ in (\ref{propagators}). Then, however, the first term of
the new $\left\langle \phi ^{\prime }~\phi ^{\prime }\right\rangle $ behaves
like $1/p^{4}$ in the infrared. This behavior generates annoying IR
divergences in $D=3$ and $D=4$. The trick (\ref{ghostv}) is safer because it
avoids this problem.

Now I\ prove that the formulation (\ref{ghostv}) admits an immediate
generalization to $SO(d)$ breaking models.

\bigskip

\textbf{The breaking of }$SO(d)$\textbf{\ to }$SO(D)\otimes SO(-\varepsilon
) $\textbf{: propagators with }$SO(d)$ \textbf{invariant denominators.} Now
I assume that the regularization preserves diffeomorphisms and the local
Lorentz symmetry group $SO(D)\otimes SO(-\varepsilon )$. The Lorentz indices 
$a,b,c\ldots $ running from $1$ to $d$ are decomposed into physical Lorentz
indices $\bar{a},\bar{b},\bar{c}\ldots $ running from $1$ to $D$, and
evanescent Lorentz indices $\hat{a},\hat{b},\hat{c}\ldots $ running from $D$
to $d$ (with $D$ excluded): $a=(\bar{a},\hat{a})$, $b=(\bar{b},\hat{b})$,
etc. The curvature tensor decomposes into 
\[
R^{\bar{a}\bar{b}}=\overline{R}^{\bar{a}\bar{b}}+\omega ^{\bar{a}\hat{c}%
}\omega ^{\bar{b}\hat{c}},\qquad R^{\bar{a}\hat{b}}=D\omega ^{\bar{a}\hat{b}%
},\qquad R^{\hat{a}\hat{b}}=\hat{R}^{\hat{a}\hat{b}}+\omega ^{\bar{c}\hat{a}%
}\omega ^{\bar{c}\hat{b}}, 
\]
etc., where $\overline{R}^{\bar{a}\bar{b}}$ and $\hat{R}^{\hat{a}\hat{b}}$
are the $SO(D)$ and $SO(-\varepsilon )$ curvatures, respectively, and $%
D_{\mu }$ denotes the $SO(D)\otimes SO(-\varepsilon )$ covariant derivative.
Observe that $\omega ^{\bar{a}\hat{b}}$ transforms as a tensor. Similarly,
objects such as $\omega _{\mu }^{\bar{a}\hat{b}}\omega _{\nu }^{\bar{a}\hat{b%
}}g^{\mu \nu }$ are scalars. The vielbein can be used to define covariant
D'Alembertians in the physical and evanescent portions of spacetime: 
\[
\overline{D^{2}}=e^{\bar{a}\mu }e^{\bar{a}\nu }D_{\mu }D_{\nu },\qquad 
\widehat{D^{2}}=e^{\hat{a}\mu }e^{\hat{a}\nu }D_{\mu }D_{\nu }. 
\]

The BRST\ transformations split as follows 
\begin{eqnarray}
se_{\mu }^{\bar{a}} &=&-e_{\rho }^{\bar{a}}\partial _{\mu }C^{\rho }-C^{\rho
}\partial _{\rho }e_{\mu }^{\bar{a}}-C^{\bar{a}\overline{b}}e_{\mu }^{\bar{b}%
},\qquad sC^{\overline{a}\bar{b}}=-C^{\bar{a}\bar{c}}C^{\bar{c}\bar{b}%
}-C^{\rho }\partial _{\rho }C^{\bar{a}\bar{b}},  \nonumber \\
s\overline{C}^{\overline{a}\bar{b}} &=&B^{\overline{a}\bar{b}}-C^{\bar{a}%
\bar{c}}\overline{C}^{\bar{c}\bar{b}}-C^{\bar{b}\bar{c}}\overline{C}^{\bar{a}%
\bar{c}}-C^{\rho }\partial _{\rho }\overline{C}^{\overline{a}\bar{b}}, 
\nonumber \\
sB^{\overline{a}\bar{b}} &=&-C^{\bar{a}\bar{c}}B^{\bar{c}\bar{b}}-C^{\bar{b}%
\bar{c}}B^{\bar{a}\bar{c}}-C^{\rho }\partial _{\rho }B^{\overline{a}\bar{b}},
\label{brs}
\end{eqnarray}
plus analogous rules obtained replacing all barred indices with hatted
indices. Here $C^{\bar{a}\bar{b}}$ are the ghosts of the physical $SO(D)$
Lorentz symmetry, $C^{\hat{a}\hat{b}}$ are the ghosts of the evanescent $%
SO(-\varepsilon )$ Lorentz symmetry and so on.

Since the symmetry $SO(d)$ is broken, it is not possible to choose a
completely symmetric gauge for the fluctuation $\widetilde{\phi }_{\mu }^{a}$%
. The simplest generalization of the symmetric gauge, 
\begin{equation}
\phi _{\bar{a}\bar{b}}^{\prime }=0,\qquad \phi _{\hat{a}\hat{b}}^{\prime }=0%
\text{,}  \label{Lgauge}
\end{equation}
kills the antisymmetric parts of the $D\times D$ and $(-\varepsilon )\times
(-\varepsilon )$ diagonal blocks of the matrix $\phi ^{\prime }$. The
components $\phi _{\bar{a}\hat{b}}^{\prime }$ are unconstrained. Since the
Einstein lagrangian (\ref{LE}) is independent of $\phi _{\bar{a}\hat{b}%
}^{\prime }$, the propagator of $\phi _{\bar{a}\hat{b}}^{\prime }$ should be
provided by extra lagrangian terms, e.g. $\sqrt{g}\omega _{\mu }^{\bar{a}%
\hat{b}}\omega _{\nu }^{\bar{a}\hat{b}}g^{\mu \nu }$. Then, however, it is
not easy to have $SO(d)$ invariant denominators. Instead, a generalization
of the Lorentz gauge functions (\ref{gaugeL}) does the job in a simple way.

The Lorentz gauge functions (\ref{gaugeL}) are replaced with the reduced set
of gauge functions 
\begin{equation}
\mathcal{G}_{rL}^{\bar{a}\bar{b}}=D^{\mu }\omega _{\mu }^{\bar{a}\bar{b}%
},\qquad \mathcal{G}_{rL}^{\hat{a}\hat{b}}=D^{\mu }\omega _{\mu }^{\hat{a}%
\hat{b}}.  \label{newgaugeL}
\end{equation}
The auxiliary fields and antighosts are correspondingly reduced to the
blocks $\overline{C}^{\bar{a}\bar{b}},B^{\bar{a}\bar{b}}$ and $\overline{C}^{%
\hat{a}\hat{b}},B^{\hat{a}\hat{b}}$, so the gauge-fixing and ghost
lagrangians in the Lorentz sector read 
\begin{eqnarray}
\widetilde{\mathcal{L}}_{\text{gf}}^{rL} &=&-\sqrt{g}\left( \frac{\overline{%
\xi }}{2}B^{\bar{a}\bar{b}}D^{2}B^{\bar{a}\bar{b}}+B^{\bar{a}\bar{b}}%
\mathcal{G}_{rL}^{\bar{a}\bar{b}}+\frac{\widehat{\xi }}{2}B^{\hat{a}\hat{b}%
}D^{2}B^{\hat{a}\hat{b}}+B^{\hat{a}\hat{b}}\mathcal{G}_{rL}^{\hat{a}\hat{b}%
}\right) ,\qquad \qquad  \label{nea} \\
\widetilde{\mathcal{L}}_{\text{ghost}}^{rL} &=&-\sqrt{g}\overline{C}^{\bar{a}%
\bar{b}}D^{\mu }\partial _{\mu }C^{\bar{a}\bar{b}}-\sqrt{g}\overline{C}^{%
\hat{a}\hat{b}}D^{\mu }\partial _{\mu }C^{\hat{a}\hat{b}},  \label{newghostv}
\end{eqnarray}
where $D^{2}=D^{\mu }D_{\mu }$. The total ghost lagrangian is 
\begin{equation}
\widetilde{\mathcal{L}}_{\text{ghost}}=\mathcal{L}_{\text{ghost}}^{\text{diff%
}}+\widetilde{\mathcal{L}}_{\text{ghost}}^{rL}.  \label{acci}
\end{equation}
The ghost propagators are still the identity times $1/p^{2}$, so their
denominators are $SO(d)$ invariant.

The sum of the Einstein lagrangian (\ref{LE}) plus the gauge-fixing terms $%
\mathcal{L}_{\text{gf}}^{\text{diff}}$ and $\widetilde{\mathcal{L}}_{\text{gf%
}}^{rL}$ of formulas (\ref{ghostu}) and (\ref{nea}), are still insufficient
to give a propagator to $\phi _{\bar{a}\hat{b}}^{\prime }$. For this
purpose, introduce an evanescent tensor field $B_{{}}^{\bar{a}\hat{b}}$
transforming as a scalar under diffeomorphisms and as a vector under both $%
SO(D)$ and $SO(-\varepsilon )$ rotations and add 
\begin{equation}
\Delta \mathcal{L}=-2\sqrt{g}\left( \frac{\xi }{2}B^{\bar{a}\hat{b}}D^{2}B^{%
\bar{a}\hat{b}}+B^{\bar{a}\hat{b}}\mathcal{G}_{rL}^{\bar{a}\hat{b}}\right)
,\qquad \text{where }\mathcal{G}_{rL}^{\bar{a}\hat{b}}=D^{\mu }\omega _{\mu
}^{\bar{a}\hat{b}},  \label{deltaL}
\end{equation}
to the Einstein action (\ref{LE}). This addition is clearly a scalar
density. In this way the matrices $B^{ab}$ and $\mathcal{G}_{rL}^{ab}$ are
fully reconstructed. The diagonal blocks $B^{\bar{a}\bar{b}},B^{\hat{a}\hat{b%
}}$ are auxiliary fields for the Lorentz gauge-fixings, while the
non-diagonal components $B^{\bar{a}\hat{b}}$ are extra evanescent fields
used for regularization. Since (\ref{deltaL}) vanishes in the formal limit $%
\varepsilon \rightarrow 0$, $\Delta \mathcal{L}$ is truly a regularization
term. Therefore, even if $B^{\bar{a}\hat{b}}$ is not BRST\ exact, its
introduction does not change the physics.

Recapitulating, the total gauge-fixed lagrangian is 
\begin{equation}
\widetilde{\mathcal{L}}_{\text{grav}}=\mathcal{L}+\Delta \mathcal{L}+%
\mathcal{L}_{\text{gf}}^{\text{diff}}+\widetilde{\mathcal{L}}_{\text{gf}%
}^{rL}.  \label{lagrali}
\end{equation}

\bigskip

Now, set for a moment the gauge-fixing parameters $\overline{\xi }$ and $%
\widehat{\xi }$ equal to $\xi $. Then the quadratic part of the lagrangian (%
\ref{lagrali}) coincides precisely with the quadratic part of $\mathcal{L}_{%
\text{grav}}$ in (\ref{lagrala}): 
\begin{equation}
\mathcal{L}_{\text{grav}}=\mathcal{L}+\mathcal{L}_{\text{gf}}^{\text{diff}}+%
\mathcal{L}_{\text{gf}}^{L}=\mathcal{L}+\Delta \mathcal{L}+\mathcal{L}_{%
\text{gf}}^{\text{diff}}+\widetilde{\mathcal{L}}_{\text{gf}}^{rL}=\widetilde{%
\mathcal{L}}_{\text{grav}}\qquad \text{for }\xi =\overline{\xi }=\widehat{%
\xi },  \label{custom}
\end{equation}
up to cubic terms, due to the different definitions of covariant
derivatives. In this case the propagators of $\phi ,\phi ^{\prime }$ and $B$
coincide with the ones of formula (\ref{propagators}).

More generally, the propagators depend linearly on $\overline{\xi }$ and $%
\widehat{\xi }$, because these are gauge-fixing parameters. It is easy to
prove by direct computation that when $\xi \neq \overline{\xi }\neq \widehat{%
\xi }$ the propagators (\ref{propagators}) are unmodified except for $%
\left\langle \phi _{\mu \nu }^{\prime }(p)~\phi _{\rho \sigma }^{\prime
}(-p)\right\rangle _{0}$, which is corrected by the replacement 
\begin{eqnarray}
-\frac{\xi }{2p^{2}}(\delta _{\mu \rho }\delta _{\nu \sigma }-\delta _{\mu
\sigma }\delta _{\nu \rho }) &\rightarrow &-\frac{\overline{\xi }}{2p^{2}}%
(\delta _{\bar{\mu}\bar{\rho}}\delta _{\bar{\nu}\bar{\sigma}}-\delta _{\bar{%
\mu}\bar{\sigma}}\delta _{\bar{\nu}\bar{\rho}})-\frac{\widehat{\xi }}{2p^{2}}%
(\delta _{\hat{\mu}\hat{\rho}}\delta _{\hat{\nu}\hat{\sigma}}-\delta _{\hat{%
\mu}\hat{\sigma}}\delta _{\hat{\nu}\hat{\rho}})+  \nonumber \\
&&-\frac{\xi }{2p^{2}}(\delta _{\bar{\mu}\bar{\rho}}\delta _{\hat{\nu}\hat{%
\sigma}}-\delta _{\bar{\mu}\bar{\sigma}}\delta _{\hat{\nu}\hat{\rho}}+\delta
_{\hat{\mu}\hat{\rho}}\delta _{\bar{\nu}\bar{\sigma}}-\delta _{\hat{\mu}\hat{%
\sigma}}\delta _{\bar{\nu}\bar{\rho}}).  \label{replad}
\end{eqnarray}
I have therefore proved that all propagators have $SO(d)$ invariant
denominators.

The regularization can be simplified if $SO(d)$ is broken just to $SO(D)$.
Then the last two terms of (\ref{nea}) are moved from $\widetilde{\mathcal{L}%
}_{\text{gf}}^{rL}$ to $\Delta \mathcal{L}$. The field $B^{\hat{a}\hat{b}}$
is interpreted as an extra evanescent tensor field, on the same footing as $%
B^{\bar{a}\hat{b}}$. The ghosts $\overline{C}^{\hat{a}\hat{b}}$, $C^{\hat{a}%
\hat{b}}$ and the last term of $\widetilde{\mathcal{L}}_{\text{ghost}}^{rL}$
in (\ref{newghostv}) are suppressed. Finally, $D_{\mu }$ is interpreted as
the $SO(D)$ covariant derivative.

Here I have used the second order formalism, but the arguments of this paper
can be easily adapted to the first order formalism, where the Lorentz
gauge-fixing (\ref{gaugeL}) looks even more natural. In the first order
formalism, gravity is described by the independent fields $e_{\mu }^{a}$ and 
$\omega _{\mu }^{ab}$. Relations (\ref{conve}) hold, except for the formula
expressing $\omega _{\mu }^{ab}$ in terms of $e_{\mu }^{a}$. The BRST\
auxiliary fields and the ghosts are unchanged. The BRST\ transformations are
(\ref{brs1}) plus 
\[
s\omega _{\mu }^{ab}=-\omega _{\rho }^{ab}\partial _{\mu }C^{\rho }-C^{\rho
}\partial _{\rho }\omega _{\mu }^{ab}-\mathcal{D}_{\mu }C^{ab}
\]
(or the corresponding reduced versions, when the continued Lorentz symmetry
is broken). The gauge-fixings and the ghost lagrangians are formally
identical: it is sufficient to interpret $\omega _{\mu }^{ab}$ as an
independent field. When the continued Lorentz symmetry is broken, the
evanescent extra tensor field(s) are the same ($B^{\bar{a}\hat{b}}$ and
eventually $B^{\hat{a}\hat{b}}$) and the evanescent deformations $\Delta 
\mathcal{L}$ are formally identical to those of the second order formalism.
However, the propagators become considerably more involved, since the
quadratic part of the lagrangian is non-diagonal in the fields $\phi _{\mu
\nu }$, $\phi _{\mu \nu }^{\prime }$, $\omega _{\mu }^{ab}$ and $B^{ab}$.
The first order formalism is useful when the torsion is non-zero and, in
particular, for the regularization of supergravity.

It is also immediate to generalize the arguments of this paper when the
metric and the vielbein are expanded around generic curved backgrounds. This
is useful for calculations with the background field method and in the
presence of a cosmological constant.

A final comment concerns the stability of the action under renormalization.
The action (\ref{lagrali}) produces nice propagators, but is obtained
choosing the evanescent deformation $\Delta \mathcal{L}$ and the
gauge-fixing in a very special way. It is natural to wonder if
renormalization spoils this structure. Divergences can be subtracted in a
RG\ invariant way only if every allowed lagrangian term is turned on,
multiplied by an independent renormalized coupling that runs appropriately.
So, to have complete RG\ invariance all types of evanescent deformations
should be included in $\Delta \mathcal{L}$ at the tree level, not just (\ref
{deltaL}), but then the propagators do not have $SO(d)$ invariant
denominators. This problem is avoided as follows. Known theorems \cite
{collins,harv} ensure that the evanescent sector of the theory does not mix
into the physical sector, but produces at most a scheme change. Therefore,
evanescent counterterms can be subtracted just as they come, at higher
orders, with no introducton of new independent parameters at the tree-level.
If a special evanescent deformation, such as the $\Delta \mathcal{L}$ of (%
\ref{lagrali}), is used at the tree level, instead of the most general
evanescent deformation, then RG invariance is violated in the renormalized
correlation functions only by contributions that vanish in the physical
limit $\varepsilon \rightarrow 0$. It is therefore possible to carry out
every calculation with the propagators produced by (\ref{lagrali}) and (\ref
{acci}). Observe that this argument is analogous to the argument commonly
used for gauge-fixing parameters: if the gauge fixing-parameter $\xi $ is
not left free to run, but set to some special value, as in the Feynman and
Landau gauges, then RG\ invariance is violated, but only in the BRST-exact
sector of the theory.

\bigskip

Summarizing, for an efficient calculation of Feynman diagrams in the
Standard Model coupled with quantum gravity using the dimensional
regularization, the gravity sector can be regularized in the way just
described and the matter sector can be regularized in the usual fashion.

The results of this paper are dimension-independent (for $D>2$). In
particular, they apply also to models such as three-dimensional Chern-Simons
gauge theories coupled with two-component fermions and gravity \cite{pap1}.
At the theoretical level, they are useful for the study of consistent
irrelevant deformations of renormalizable theories and the predictivity of
certain classes of power-counting non-renormalizable theories, such as those
studied in ref.s \cite{pap2,pap3}. At the phenomenological level, they are
useful for calculations of gravitational radiative corrections in low-energy
phenomenological models.

\end{document}